\documentclass[12pt]{article}
\begin{document}
\newcommand{\la} {\langle}
\newcommand{\ra} {\rangle}

%This is the double spaced hard copy .

\title{Resource Letter TF-1: Turbulence in Fluids}

\author{Mark Nelkin$^{1}$ } 

\maketitle

\noindent $^{1}$Physics Department, New York University, New York, NY 10003,\\
and Levich Institute, CCNY, New York, NY 10031 USA\\
Electronic address: Mark.Nelkin@nyu.edu

\renewcommand{\baselinestretch}{2}  

\begin{abstract}
This Resource Letter provides a guide to the literature on fully developed turbulence in fluids.  It is restricted to mechanically driven turbulence in an incompressible fluid described by the Navier-Stokes equations of hydrodynamics, and places greatest emphasis on fundamental physical questions.  Journal articles and books are cited for the following topics: The Navier-Stokes equations, qualitative aspects of turbulence, the 1941 Kolmogorov theory, intermittency and small scale structure, time correlations and pressure; with brief mention of  two-dimensional turbulence, passive scalars in turbulence, and the turbulent boundary layer,

\end{abstract}

\pagebreak

\setlength{\baselineskip}{20pt}

Turbulent fluid flows are ubiquitous in the atmosphere, the oceans and the stars.  They also occur in a wide variety of engineering applications.  Most studies of turbulence have an applied objective, whether  this application be to engineering, to geophysics, to astrophysics or to weather predicition.  But there is a basic problem in physics underlying all of these applications.  This Resource Letter provides a guide to the literature relevant to this problem.

\section*{THE NAVIER-STOKES EQUATIONS}

Consider a restricted class of fluid flows in which the variations of fluid density are negligible.  This excludes the important subject of buoyant convection.  It also excludes high Mach number flows where the dynamic effects of pressure become important.  Assume that the fluid satisfies Newton's law of viscosity.  With this restriction we are left with incompressible fluid flow governed by the Navier-Stokes equations of hydrodynamics.  These are partial differential equations for a velocity field 
${\bf v(r},t)$, where $\bf r$ is the position vector and $t$ is time. Because the density is assumed constant, this field is divergenceless.

\begin{equation}
{\bf\nabla}\cdot{\bf v(r},t)=0.
\label{div}
\end{equation}
The momentum balance of a moving fluid element is described by the Navier-Stokes equations

\begin{equation}
\partial {\bf v}/\partial t + ({\bf v}\cdot{\bf \nabla}) {\bf v}= -(1/\rho){\bf \nabla}p+\nu \nabla^2 {\bf v},
\label{navier}
\end{equation}
where $p({\bf r},t)$ is the dynamic pressure field, $\rho$ is the constant density, and $\nu$ is the kinematic viscosity of the fluid.  When supplemented with the ``stick" boundary condition that the relative velocity of the fluid with respect to any bounding surface is zero, Equations (1) and (2) give a complete mathematical description of the dynamics of incompressible fluid flow. These equations should apply accurately whenever the density changes are small, and the flow is slowly varying on a molecular space and time scale; conditions that are comfortably met in most air and water flows on a laboratory or geophysical scale.

The kinematic viscosity $\nu$ is the only molecular property of the fluid that enters the equations.  If the density is constant, this is the only way to distinguish fluid flows in air from fluid flows in water.  Suppose we have a flow with a geometrically well-defined largest length scale $L$, and a well-defined largest velocity scale $U$.  When put in dimensionless form, the Navier-Stokes equations contain only one dimensionless parameter, the Reynolds number, $Re$, defined by

\begin{equation}
Re=UL/\nu.
\end{equation}
All incompressible flows with the same Reynolds number and the same flow geometry should have the same flow properties when measured in the appropriate units.  This $Re$ scaling applies when $Re$ is small, and the flow is laminar and well understood.  It applies equally well when $Re$ is large, and the flow is turbulent and poorly understood. The statistical properties of turbulence are the same in air and water flows when the Reynolds number is the same, and the flow geometry is the same.  The only thing that air and water have in common is the Navier-Stokes equations.  Thus we have confidence that we are starting from the right equations.  

There are a large number of books on fluid mechanics.  I mention here only four that are particularly appropriate for physicists.  The book by Faber \cite{faber} is a good introduction.  The book by Tritton \cite{tritton} emphasizes the role of experiment, and the book by Landau and Lifshitz \cite{landau} remains a classic introduction to fluid mechanics as a part of theoretical physics. The book by Batchelor \cite{batchelor70} gives a systematic theoretical introduction to the subject.

\section*{QUALITATIVE ASPECTS OF TURBULENCE}

Qualitatively, turbulence is not a mystery.  The Navier-Stokes equations are a dissipative dynamical system with many degrees of freedom, and the effective number of degrees of freedom increases strongly with increasing $Re$.  For low $Re$, we observe and can compute smooth laminar solutions with little spatial structure.  As $Re$ increases, the generic sequence of events is a transition to a steady spatially-structured flow, then to a spatially-structured and time-periodic flow, and then to a flow that is chaotic in time.  This behavior is qualitatively similar for different flow geometries, but is quantitatively far from universal.  There are many recent books on chaos and the transition to turbulence.  I will not discuss this subject in this Resource Letter, but a clear presentation at an accessible level is given by Berg\'e {\it et al} \cite{berge}.

I will consider only fully developed turbulence, which occurs for $Re$ much larger than the critical Reynolds number $Re_c$, where chaos first occurs.  High $Re$ turbulence is chaotic in both space and time, but also exhibits considerable spatial structure.  The velocity fluctuations occur on a wide range of space and time scales.  The large-scale fluctuations, often called the large eddies, have a size of the order of the geometrically imposed external length scale $L$.  These large eddies are not universal with respect to the flow geometry, and are of dominant importance in the energetics and transport properties of the flow.  A good general introduction to this aspect of turbulence is given by Tennekes and Lumley \cite{tennekes}.  A more detailed description from an engineering perspective is given by Hinze \cite{hinze}.  A systematic development of the statistical theory of turbulence is given by Lesieur \cite{lesieur}. An excellent review, from a qualitative and historical perspective, has recently been given by Sreenivasan \cite{sreeni99}.

The small-scale fluctuations, with sizes $r<<L$, have universal statistics independent of the flow geometry, and are of great interest for fundamental physics.  The dominant strategy for computing and understanding of turbulent flows is large-eddy simulation (LES).  The universal small scales are modeled, and this model is used for explicit computation of the large-scale fluctuations.  I will not discuss LES in this Resource Letter, but a good review can be found in \cite{les}.

Direct numerical simulation (DNS) of the Navier-Stokes equations is limited to modest values of $Re$.  Suppose the smallest eddies in a turbulent flow are of size $\eta$.  The number of coupled ordinary nonlinear differential equations that must be solved is of order $(L/\eta)^3$.  Experiment and phenomenological theory to be discussed below indicate that $(L/\eta)$ scales as $Re^{3/4}$.  Thus there are about $10^6$ degrees of freedom in a turbulent flow with $Re=1000$.  But the values of $Re$ in typical laboratory turbulent flows are about $10^4$ to $10^5$, and in the atmosphere of order $10^6$ to $10^7$.  DNS plays an increasingly important role, particularly in comparison with relatively low $Re$ laboratory flows, but it is not a substitute for a physical understanding of the statistical properties of turbulence, nor is it a possible strategy for the direct computation of high $Re$ turbulent flows.  A good recent review of DNS by Moin and Mahesh is available \cite{dns}.

The mathematics of the Navier-Stokes equations poses many fundamental problems.  The overall questions of existence, uniqueness, and smoothness of solutions are largely unresolved.  I am not qualified to discuss this subject in detail, but I mention only one example.  It is not known if the Euler equations, which are the Navier-Stokes equations when the viscosity $\nu=0$, develop spatial singularities in a finite time when starting from smooth initial conditions.  
An interesting review of some of the mathematical subtleties of this problem has been given by
Constantin \cite{constantin94}.
At a more applied level, there is an increasing interaction between the mathematics of the Navier-Stokes equations and the theory of turbulence.  This is particularly true for the calculation of variational bounds.  An excellent introduction to the mathematics of the Navier-Stokes equations is given in the book by Doering and Gibbon \cite{doering95}.  Two recent articles \cite{doering94} \cite{grossmann97} give a flavor of recent work.

\section*{THE 1941 KOLMOGOROV THEORY}

The one striking success in turbulence theory is the phenomenological picture introduced by Kolmogorov in 1941, which has been critically analyzed and carefully explained in the recent book by Frisch \cite{frisch}.  The essential qualitative idea is a cascade of energy from large scales to small scales with an eventual dissipation of energy at the smallest scales.  The decreasing time scales during the cascade of energy lead to a statistically steady state.  In this steady state there is one key dynamical quantity. This is $\la \epsilon \ra$, the average rate of energy dissipation per unit mass. The quantity 
$\la \epsilon \ra$ plays multiple roles.  It is the rate at which energy is fed into the turbulent fluctuations at large scales, it is the rate at which energy is transferred from large to small scales by the nonlinear terms in the Navier-Stokes equations, and it is also the rate at which the energy is dissipated at the smallest scales by the action of molecular viscosity.  

The Kolmogorov school of turbulence research in Russia had a profound influence on the development of the subject.  I do not give explicit references to the original Russian literature from the 1940s and 1950s.  This is done in the book by Frisch \cite{frisch}, and more completely in the monumental and still very useful book by Monin and Yaglom \cite{monin}.  The history of Kolmogorov's school has been reviewed in a delightful article by Yaglom \cite{yaglom94}.

An essential feature of the 1941 Kolmogorov picture is the statistical universality of the small-scale fluctuations.  The basic dynamical coupling is assumed to be between scales of similar size so that the direct coupling of the largest and smallest scales is small.  
A key assumption is that $\la \epsilon \ra$ is independent of viscosity.  The evidence supporting this assumption from DNS has been reviewed by Sreenivasan \cite{sreeni}. In the 1941 Kolmogorov theory, $\la \epsilon \ra$ is the only relevant dynamical property of the flow.  When combined with $\nu$, the kinematic viscosity of the fluid, the smallest length scale in the flow can be calculated by dimensional analysis, and is given by

\begin{equation}
\eta=(\nu^3/\la \epsilon \ra)^{1/4} \sim L Re^{-3/4}.
\label{eta}
\end{equation}
It should be emphasized that (\ref{eta}) is a prediction in terms of directly measurable quantities, as discussed, for example, in the review article by Nelkin \cite{nelkin94}.

To be more quantitative, consider the basic correlation function

\begin{equation}
B_{LL}(r,t) = \la u(x,y,z,t) u(x+r,y,z,t) \ra,
\label{ucorr}
\end{equation}
where ${\bf r}=(x,y,z)$ and the velocity field ${\bf v}=(u,v,w)$, and the averaging $\la  \;\; \ra$ is  a time average over a statistically stationary flow.  The most frequently studied quantity is the longitudinal energy spectrum $E_{11}(k)$, which is the Fourier cosine transform of (\ref{ucorr}).  In the 1941 Kolmogorov picture, the only relevant dynamical parameters are $\la \epsilon \ra$ and $\nu$.  Dimensional analysis then requires that $E_{11}(k)/(\la \epsilon \ra \nu^5)^{1/4}$ be a universal function of $k\eta$, where $\eta$ is calculated from (\ref{eta}). This means that the scaled energy spectrum should fall on a single universal curve for a wide variety of experiments.  This works extremely well, as illustrated by Figure 9 of the paper by Saddoughi and Veeravalli \cite{saddoughi}.

The intermediate wave-number region,

\begin{equation}
(1/L)<<k<<(1/\eta),
\label{range}
\end{equation}
is known as the inertial sub-range.  In this wave-number range, statistical universality is expected, but the molecular viscosity $\nu$ should play no dynamical role.  In this range, dimensional analysis implies  that 

\begin{equation}
E_{11}(k)= C_1 \la \epsilon \ra^{2/3} k^{-5/3},
\label{inertial}
\end{equation}
where $C_1$ is called the Kolmogorov constant.  Sreenivasan \cite{sreeni95} has reviewed the experimental evidence for the universality of the Kolmogorov constant, and finds that all of the high Reynolds number data are consistent with a value of $C_1\approx 0.5$.  Yeung and Zhou \cite{yeung97} have reviewed the evidence from DNS with essentially the same conclusion.

A fundamental assumption of the 1941 Kolmogorov theory is that the small-scale fluctuations are statistically isotropic. The consequences of this assumption at the level of second-order statistics are fairly simple, and are discussed in detail in Monin and Yaglom \cite{monin}.  For example, the transverse energy spectrum $E_{22}(k)$ in the power-law range of (\ref{inertial}) is just $(4/3)$ times the longitudinal spectrum $E_{11}(k)$.  A more sensitive test is to examine a quantity that would be identically zero for isotropic turbulence.  The simplest such quantity is the cross-correlation function 
$\la u(x) v(x+r) \ra$, whose fourier transform $E_{12}(k)$ is known as the cospectrum.  A simple dimensional argument first given by Lumley \cite{lumley67} suggests that this should decay as 
$k^{-7/3}$ so that anisotropy becomes smaller as the wave number becomes larger.  This seems reasonably well supported by experiment \cite{saddoughi}.

\section*{INTERMITTENCY }

In condensed-matter physics, particularly in critical phenomena, ideas of universality and scaling play an important role, but the observable statistical properties are typically limited to a few low-order correlation functions.  In turbulent fluid flow the velocity signal $u(x)$ is observable as a random process.  It has long been known that the random process describing the turbulent velocity field is not Gaussian, and that important information can be obtained by examining higher-order statistical quantities.  In particular, the turbulent signal tends to occur in intermittent bursts, and this tendency is increasingly strong as the scale of the motion is decreased.   A good introduction to the idea of intermittent random processes as it applies in turbulence is given in the book by Frisch \cite{frisch}.  Our knowledge of these higher-order statistics has been summarized in the review articles by Nelkin \cite{nelkin94} and by Sreenivasan and Antonia \cite{sreeni97}.

One of the earliest confirmations of the Kolmogorov $5/3$ law (\ref{inertial}) for the energy spectrum \cite{grant62} also showed that the dissipation exhibits very strong fluctuations.  
In fact, the most important defect in the 1941 Kolmogorov picture is probably the replacement of the fluctuating local dissipation by its average. 
Experiments are usually limited to a one-dimensional surrogate for the local dissipation,

\begin{equation}
\epsilon(x)=15 \nu (\partial u/\partial x)^2. 
\end{equation}
This local dissipation exhibits anomalous 
scaling,  most naturally characterized by power-law behavior of its autocorrelation function $\la \epsilon(x) \epsilon(x+r)\ra$, when $r$ is an inertial range distance:

\begin{equation}
\la \epsilon(x) \epsilon(x+r)\ra = Const. \la \epsilon \ra^2 (L/r)^{\mu},
\end{equation}
where $\mu$ is a putatively universal exponent describing the dissipation fluctuations.  The experimental evidence on the intermittency exponent $\mu$ has been reviewed by Sreenivasan and Kailasnath \cite{sreeni93}. They find that all of the data are consistent with a value of $\mu=0.25$, with an uncertainty of about $20\%$.

Most studies of anomalous scaling in the inertial range have concentrated on the longitudinal structure functions,

\begin{equation}
D_p(r)=\la [u(x+r)-u(x)]^p \ra = \la [\Delta u(r)]^p \ra,
\end{equation}
which are the moments of the velocity difference $\Delta u(r)$ between two points separated by a distance $r$.  The term ``longitudinal" refers to the component of the velocity difference along the line joining the two points.  These structure functions are characterized by power-law behavior in the inertial range with anomalous scaling exponents $\zeta_p$ defined by
\begin{equation}
D_p(r)=Const. \la \epsilon \ra^{p/3} \; r^{p/3} (r/L)^{(\zeta_p-p/3)}.
\label{zeta}
\end{equation}
In the 1941 Kolmogorov picture, the velocity distribution is self-similar, and the exponents $\zeta_p$ take the ``normal" values of (p/3).  As discussed in \cite{frisch} and \cite{monin}, an exact theoretical result from Kolmogorov requires that $\zeta_3=1$.  The appearance of the external length scale $L$ in (\ref{zeta}) indicates that the physics of anomalous scaling is associated with the buildup of fluctuations as the cascade process proceeds to smaller scales.

The experimental values of the scaling exponents $\zeta_p$ are quite robust, with agreement among many experiments, and also with DNS.  See for example, Fig. 3 of \cite{sreeni97} for experiment, and table 1 of \cite{cao96} for DNS. It is remarkable that these universal scaling exponents are observed even in superfluid helium \cite{maurer98}, but it is not well understood why superfluid and classical turbulence should be so nearly the same when the superfluid is flowing rapidly.  In obtaining these scaling exponents for a limited inertial range, most authors use the extended self-similarity idea due to Benzi {\it et al} \cite{benzi93} \cite{benzi96}.  The basic idea is to extend the scaling range by plotting the p'th order structure function $D_p(r)$ {\it versus} $D_3(r)$ instead of {\it versus} $r$.  This gives a greatly extended power-law range at finite Reynolds number, but makes little difference at high Reynolds number.  It is not understood theoretically why this procedure works so well.

A good fit to the experimental and DNS values of $\zeta_p$ is given by the model  proposed by She and Leveque \cite{she94}

\begin{equation}
\zeta_p=(p/9)+2[1-(2/3)^{p/3}].
\label{sheleveque}
\end{equation}
One feature of the She-Leveque model is its prediction in the limit of large $p$.  It was suggested by Novikov \cite{novikov94}, using arguments from probability theory, that the limit of $\zeta_p/p$ for large $p$ should be zero.  It was pointed out by Nelkin \cite{nelkin95} that there is no possibility of settling this question from experiment, but it remains an interesting theoretical question.

As first suggested by Kolmogorov in 1962 \cite{k62}, the anomalous scaling of the dissipation is closely related to the anomalous scaling of the velocity field.  This is known as Kolmogorov's refined-similarity hypothesis (RSH), and is usually stated in the form

\begin{equation}
D_p(r) = Const \la \epsilon_r^{p/3} \ra r^{p/3},
\label{rsh}
\end{equation}
where $\epsilon_r$ is the average of the local dissipation over an inertial-range interval of size $r$.  As pointed out by Nelkin and Bell in 1978 \cite{nelkin78}, this hypothesis reflects the reasonable physical assumption that the locally averaged dissipation should have the same statistical properties as the locally averaged nonlinear energy transfer, and that this latter quantity is reasonably represented by
$[\Delta u(r)]^3/r$.  In recent years, the RSH has developed increasing experimental support 
\cite{stolo92} \cite{praskov97}. The RSH can also be profitably studied in the more general context of stochastic processes \cite{stolo94}.

There is also considerable interest in the Reynolds number dependence of single-point quantities such as the velocity-derivative flatness

\begin{equation}
F=\la (\partial u/\partial x)^4 \ra / \la \partial u/\partial x)^2 \ra^2= 
Const. \la {\epsilon}^2 \ra / \la \epsilon \ra^2.
\end{equation}
Experimentally, this is typically much larger than the Gaussian value of $3$.  A reasonable scaling assumption suggests that the flatness will increase as a power of the Reynolds number, reflecting the tendency of the dissipation to become increasingly intermittent as the Reynolds number increases.  This is supported by most of the experimental data, as reviewed in \cite{sreeni97}, but there is one interesting and controversial exception.  Experiments by Tabeling's group in Paris \cite{belin97},  in the flow between counter-rotating disks at high Reynolds number, done in helium gas near its critical temperature, suggest that the flatness exhibits a plateau at intermediate Reynolds number, perhaps to start increasing again at higher Reynolds number.

The observation that the moments of  $\Delta u(r)$ scale as power laws in $r$, but with independent nontrivial scaling exponents $\zeta_p$ is called multiscaling.  This can be given a geometric interpretation in terms of multifractals, as first introduced in turbulence by Mandelbrot in 1974 \cite{mandelbrot74}.  This idea has been extensively developed in the context of the local dissipation by Meneveau and Sreenivasan \cite{meneveau91}, and is reviewed for the velocity field in the book by Frisch \cite{frisch}.

The multiscaling behavior of the velocity statistics naturally suggests a description in terms of multiplicative random processes.  This was started in the 1960s by the Kolmogorov school, and is described in \cite{monin}.  Its most elegant formulation was given by Novikov \cite{novikov71} \cite{novikov90}. There are some problems in applying this description to the turbulent dissipation \cite{nelkin96}, but these problems can be avoided by careful attention to certain artifacts of the observation procedure \cite{jouault98}.  Recently, the idea of an uncorrelated multiplicative random process has been applied directly to multiscale velocity correlations \cite{benzi98}, and seems to work extremely well.

It is tempting to associate the anomalous scaling described above with the presence of coherent small-scale structures in the flow.  Since the early DNS of Siggia \cite{siggiadns} and Kerr \cite{kerrdns}, it has been known that vortex filaments are present in isotropic turbulence, at least at moderate Reynolds number.  These filaments appear very clearly in the DNS by She, Jackson and Orszag \cite{shedns}. They appear to have lengths of the order of  $L$, and diameters of the order of the dissipation scale $\eta$.  These filaments play an important role in the model of She and Leveque (\ref{sheleveque}) for the exponents $\zeta_p$.  The relevance of these vortex filaments to the dynamics of intermittency remains, however, controversial.  In particular, Jimenez {\it et al} \cite{jimenez93} have raised the issue of the stability of these vortex filaments at high Reynolds number.

To understand intermittency and small-scale structure directly from Navier-Stokes dynamics remains a formidable theoretical challenge.  The essential physics was first described by Kraichnan in 1974 \cite{rhk74}.  The nonlinear $({\bf v}\cdot{\bf \nabla}) {\bf v}$ term in the Navier-Stokes equations tends to form shocks and enhance intermittency while the pressure gradient term leads to long-range spatial coupling that depresses intermittency.  The goal of dynamical theory is to calculate the delicate competition between these two terms.  An interesting approximate theory based on this idea has been given recently by Yakhot \cite{vy98}.  An alternative point of view starts from statistical field theory.  There is a large literature developing in this area, but it is difficult, and most of it has not yet demonstrated applicability to real turbulent flows.  One direction in which some success has been achieved is  to obtain a more fundamental understanding of relations between different scaling exponents, as given for example by the Kolmogorov refined-similarity hypothesis.  There is a large effort by the group at the Weizmann institute in this area.  One of their papers \cite{lvov97} is of particular interest in this connection.

Finally, there has been considerable recent interest in the comparative scaling of longitudinal and transverse components of the velocity field.  At the level of single-point averages involving velocity fields, Siggia \cite{siggia81} showed that there are four independent invariant quantities in isotropic tubulence that can be formed from products of four elements of the velocity-derivative tensor.  Thus there are four independent velocity-derivative flatness quantities.  Two of these are the mean-square dissipation and the mean-square enstrophy.  The term ``enstrophy" was introduced by Kraichnan for the squared vorticity, and the term has remained in common use.  The simplest question one can ask is whether the mean-square dissipation and mean-square enstrophy scale in the same way with Reynolds number.  The indications from early DNS \cite{siggiadns} \cite{kerrdns} are that they do not, with the enstrophy being more intermittent than the dissipation.  Recently, attention has focused on the inertial-range scaling of the averaged dissipation $\epsilon_r$ and the corresponding enstrophy quantity.  Recent DNS shows once again \cite{enstrdns97} that the enstrophy is more intermittent than the dissipation at moderate Reynolds number.  There are two more recent papers, however, that strongly suggest that this can not persist in the high Reynolds number limit.  A group at Los Alamos \cite{he98}, starting from an arbitrary collection of static cylindrical vortices, has shown that such vortices must have the same asymptotic scaling of dissipation and enstrophy in the high Reynolds number limit even though apparent differences in scaling can persist to quite high Reynolds numbers.  A simple argument starting from the Poisson equation for the pressure \cite{nelkin99} reaches the same conclusion about the asymptotic behavior on quite general grounds in only a few lines of algebra.

The situation with respect to the transverse velocity-structure functions $\la[\Delta v(r)]^p\ra$ remains unclear.  There are experiments and DNS that show an apparent difference in scaling exponents between longitudinal and transverse structure functions, but it is not clear if these differences are real, or are an artifact of a limited scaling range \cite{sreeni98}.  Attempts to extract the effects of anisotropy in scaling are still at an early stage \cite{arad98}.

\section*{TIME CORRELATIONS AND PRESSURE}

The discussion so far has considered only equal-time correlation functions.  In equilibrium statistical physics, time correlations contain important additional dynamical information.  In turbulence, to first approximation, they do not.  In fact, the equal-time correlation functions discussed so far are usually measured with a single probe that gives the velocity $u(t)$ {\it versus} time at a fixed spatial point.  Then Taylor's frozen-turbulence assumption is invoked.  This states that the small scales are swept by the point of observation by the mean flow without dynamical distortion so that time can be converted into space through $r=Ut$, where $U$ is the local mean speed.  Corrections to the Taylor approximation are subtle and difficult.  I give only one recent reference \cite{dahm97}.  At a theoretical level, this ``sweeping" effect is largely responsible for the difficulty in developing a statistical theory of turbulence.  A qualitative discussion is given by Frisch \cite{frisch}.

The above discussion applies to an Eulerian framework, sitting at a fixed spatial point.  It is also possible to take a Lagrangian viewpoint, moving with a fluid element.  This has important applications to pollutant dispersion and to chemical reactions, and there has been some good DNS work, particularly by Yeung and collaborators \cite{yeung99}.  Until recently, Lagrangian measurements in the laboratory have not had adequate resolution to resolve the smallest scale motions.  This has now changed because of new detection techniques, and the first results are available \cite{voth98}.  It is instructive to focus on the local fluid-particle acceleration, which is just the right-hand side of the Navier-Stokes equation (\ref{navier}).  As discussed in \cite{monin}, this is dominated by the pressure term so that the statistics of particle acceleration are essentially those of the pressure gradient.  We will return to this after a more general discussion of the statistics of pressure.

Taking the divergence of the Navier-Stokes equation, and using the incompressibility condition (\ref{div}), the pressure field satisfies the Poisson equation,

\begin{equation}
		\nabla^2 p = -(\partial v_i/\partial x_j) (\partial v_j/\partial
x_i),   
\label{poisson}
\end{equation}
where the density $\rho$ is taken as one.  Thus the statistics of the pressure are determined by the statistics of the velocity field at the same time, and do not contain independent information. 
Small-scale pressure measurements in a turbulent flow are very difficult so that the problem has not received much attention, but there are some useful theoretical results.  Dimensional analysis of the 1941 Kolmogorov type is discussed in \cite{monin} as is a quasi-Gaussian approximation that allows the pressure spectrum to be expressed in terms of the energy spectrum $E_{11}(k)$.  The expected inertial range scaling for the pressure is $\la \epsilon \ra^{4/3} k^{-7/3}$, but this is not observed in DNS \cite{yeung99} at modest Reynolds number.

An important theoretical contribution to understanding the pressure statistics in isotropic turbulence was made by Hill and Wilczak \cite{hill95}.  They derived an exact expression for the pressure
 structure function $\la \Delta p(r)^2 \ra$ in terms of the three independent fourth-order velocity structure functions $\la \Delta u(r)^4 \ra$, $\la \Delta v(r)^4 \ra$ and $\la \Delta u(r)^2 \Delta v(r)^2\ra$. ( See \cite{monin}, Equation (13.82) for an exact expression of the most general fourth-order velocity difference in terms of these three independent quantities.)  When this expression is applied using the fourth-order velocity structure functions from DNS \cite{nelkin98}, it is observed that the pressure structure function is extremely sensitive to small apparent differences in scaling among  these quantities.  Thus it is not surprising that the 1941 Kolmogorov theory does not work well for the pressure at modest Reynolds number.  

Finally there is an interesting controversy concerning the scaling of the mean square particle acceleration.  The Lagrangian DNS of Vedula and Yeung \cite{yeung99} and of Gotoh and Rogallo \cite{gotoh99} show that this quantity increases much more rapidly with Reynolds number than suggested by 1941 Kolmogorov scaling.  The experiments at higher Reynolds number, however, \cite{voth98} indicate that the Reynolds number dependence is consistent with 1941 Kolmogorov scaling.  Recently, Nelkin and Chen \cite{nelkin99b} have suggested a possible resolution of this controversy, indicating that the DNS are strongly influenced by low Reynolds number corrections which do not extrapolate to the higher Reynolds numbers of the experiments, and that the experimental value of the mean square particle acceleration is numerically much too large to be accounted for by the 1941 Kolmogorov theory.

\section*{TWO-DIMENSIONAL TURBULENCE}

The principal dynamical mechanism in turbulence is the stretching of vorticity.  In two dimensions, the vorticity is perpendicular to the velocity so that this mechanism is absent, and turbulence in the usual sense can not exist.  In a seminal paper in 1967, Kraichnan \cite{rhk67} pointed out, however, that important and interesting phenomena, related to turbulence, could occur in two dimensions.  Although a cascade of energy from large scales to small scales is not possible, there is the possibility of an ``inverse cascade" of energy from small scales to large, and there is also the possibility of a cascade of enstrophy from large scales to small.  The enstrophy cascade was also proposed by Batchelor in 1969 \cite{batchelor69}.  A good critical review of earlier work is given by Frisch \cite{frisch}.

In three dimensions there is no essential difference between freely decaying and forced turbulence.  Since the cascade is accelerating, a quasi-steady state is reached in the decaying case, and a steady state in the forced case.  In two dimensions, the cascade regimes are much less robust.  Consider first the inverse cascade of energy in a system forced at wave number $k_0$.  Energy flows to smaller $k$, but the time scales increase as the wave number decreases.  In an infinite system, a steady state is never reached, but the energy flows to smaller wave numbers as time increases.  In a finite system, the inverse cascade regime is a transient effect, as clearly seen in the DNS of Smith and Yakhot \cite{smith94}.  When the largest scales become comparable to the size of the computational box, energy is broadly redistributed, and the cascade regime is terminated.  To obtain a steady inverse cascade, some additional damping mechanism is needed.  This has been achieved in the laboratory by Paret and Tabeling \cite{paret97} where a steady $k^{-5/3}$ regime is clearly observed.  The damping in this case comes from a parabolic velocity profile in the third dimension.  In this same experimental configuration, Paret and Tabeling \cite{paret98} observe that the probability distribution of  $\Delta u(r)$ is nearly Gaussian, and that anomalous scaling of the exponents $\zeta_p$ is absent.  This is an important result that deserves further theoretical study.

The enstrophy cascade is proposed to have a $k^{-3}$ energy spectrum, but this is not a robust scaling regime and can be affected by many conditons of the flow.  For a DNS of decaying two-dimensional turbulence, see Chasnow \cite{chasnow97}.  For a review of earlier DNS see Frisch \cite{frisch}.  There have been several interesting laboratory experiments recently on two-dimensional turbulence in soap films, started by Goldburg's group at Pittsburgh.  The most recent of their experiments \cite{goldburg98} gives a good confirmation of the $k^{-3}$ spectrum for decaying turbulence. A recent experiment by Rivera {\it et al} \cite{rivera98} critically analyzes the effects of compressibility due to changes in film thickness, and confirms the quasi two-dimensional nature of the flow.  An experiment in a soap film with forcing \cite{rutgers98} exhibits a steady state with a simultaneous direct enstrophy cascade and inverse energy cascade.  Finally, Hansen, Marteau and  Tabeling \cite{tabeling98} have studied the decay of unforced two-dimensional turbulence in their electromagnetically driven system.  This subject has been greatly stimulated by recent experiments that  hopefully will lead to more sophisticated theoretical understanding, and to more relevant direct numerical simulations.

\section*{PASSIVE SCALARS IN TURBULENCE}

A passive scalar is a diffusive contaminant in a fluid flow that is present in such low concentration that it has no dynamical effect on the fluid motion itself.  Examples are the temperature when the heating is weak enough that buoyancy can be neglected, and the concentration of moisture in air or dye in water.  Passive scalars have important applications, but I will discuss here only recent developments related to basic physics.  An excellent review of the subject by Warhaft \cite{warhaft2000} has been published  recently. There are two subjects where recent fundamental developments have occurred.  The first is the absence of small-scale isotropy for the scalar. The second is the intermittency of the passive scalar, which is strong in a turbulent flow, but which is present even when the velocity field is Gaussian and exhibits no intermittency.

Suppose that the passive scalar concentration is $\Theta({\bf r},t)$. If the scalar field is isotropic, the scalar-derivative skewness

\begin{equation}
S_{\Theta}= \la (\partial \Theta/\partial x)^3 \ra /[ \la (\partial \Theta/\partial x)^2 \ra]^{3/2}
\label{skew}
\end{equation}
must be identically zero.  Since the derivative $(\partial \Theta/\partial x)$ is a small-scale quantity, this quantity should be small if the scalar is locally isotropic, {\it i.e.}, is isotropic at small scales.  This was expected by analogy to the velocity field, where small-scale isotropy is reasonably well satisfied.  It has been known for a long time, however, that the scalar-derivative skewness is of order unity in high Reynolds number flows.  The experimental situation has been reviewed by Sreenivasan \cite{sreeni91}. More recently it has become clear that scalar anisotropy depends only on the presence of a mean scalar gradient, and occurs even for a Gaussian velocity field.  This was first shown by numerical simulation by Holzer and Siggia \cite{holzer}. Tong and Warhaft \cite{tong} showed that the scalar-derivative skewness was substantial in isotropic turbulence with a mean temperature gradient.  These results have been extended by Mydlarski and Warhaft \cite{mydlarski98a}. This work was motivated by the theoretical studies by Shraiman and Siggia \cite{shraiman96} that showed that small-scale scalar anisotropy was to be expected on general theoretical grounds.  This has been elaborated further and extended by theoretical and experimental study of three-point scalar correlations \cite{mydlarski98b} \cite{mydlarski98c}.  Further references can be found in the review by Warhaft \cite{warhaft2000}.

It has long been known \cite{sreeni97} \cite{warhaft2000} \cite{mydlarski98a} that the scaling exponents defined by the inertial-range scalar difference $\la [\Delta \Theta (r)]^p\ra$ show greater intermittency than the corresponding exponents for the velocity field.  This intermittency depends quantitatively on the velocity field, but an important development has been to realize and understand that the scalar will exhibit intermittency even for a Gaussian velocity field.  Theoretical activity in this subject has been intense, starting with the introduction of a model system by Kraichnan \cite{rhk68}, and rapidly intensifying when Kraichnan showed that this model system gave scalar intermittency \cite{rhk94}. The model system is a Gaussian velocity field with a velocity structure function varying as a power of the separation $r$ at all scales $r$, but varying very rapidly in time.  Kraichnan showed conclusively \cite{rhk94} that this system gave anomalous scaling of the passive scalar, and gave an approximate analytic expression for the scaling exponents.  Direct numerical simulation of this artificial problem is possible, but it is difficult to get a good scaling range in the usual Eulerian picture \cite{chen98}.  Recently, however, two groups discovered independently \cite{frisch98} \cite{gat98} that this problem could be studied from a Lagrangian point of view, following the motion of scalar particles, and that this Lagrangian picture was immensely more efficient numerically for the calculation of anomalous exponents.  There has been intense theoretical activity in calculating these anomalous exponents from perturbation theory in various limits.  This is a difficult theoretical literature that is too far afield to review here, but see \cite{frisch98} and \cite{gat98} for references.

The passive scalar problem has led to a lot of new physics.  It is not known how much of this will help us to better understand the turbulent velocity field.  Qualitatively the scalar shows much less universality than the velocity field.  This is not surprising to me since I would not expect universality for this linear problem.  

\section*{THE TURBULENT BOUNDARY LAYER}

In real turbulent flows, turbulence is usually generated at solid boundaries.  At high enough Reynolds numbers, the bulk of the flow will not show strong effects of the boundaries, and the 1941 Kolmogorov ideas of locally isotropic turbulence apply.  Near the boundaries, shear effects are important and must be explicitly considered.  As a first question, consider under what conditions boundary effects can be neglected in determining the average properties of the flow.  A careful experiment was carried out in Austin, Texas, \cite{lathrop92} on the Reynolds number dependence of the torque for circular Couette flow up to $Re=10^6$.  Circular Couette flow is the flow between concentric counter-rotating cylinders.  The prediction of 1941 Kolmogorov scaling is that the torque should go as $Re^2$, and this should be an upper bound according to the variational theories previously cited \cite{doering94} \cite{grossmann97}.  Experimentally, the torque never reaches this upper bound, and the effective exponent in torque versus $Re$ never reaches its supposed asymptotic value of $2$.  This is presumably due to persistent effects of the boundary layer.  This has been elegantly demonstrated by  a group at ENS Paris \cite{cadot97}.  They measured the overall energy dissipation and the torque for circular Couette flow with and without small vanes attached to the cylinders to break up the boundary layer.  Without the vanes they obtained results in agreement with the Texas group \cite{lathrop92}.  With the vanes, they obtained 1941 Kolmogorov scaling to a good approximation both for the torque and for the energy dissipation.  They performed a similar experiment on the flow between counter-rotating disks, and obtained similar results.  Thus the idea that 1941 Kolmogorov scaling applies to the bulk of the flow, but with significant corrections due to shear near boundaries seems well established.  This still leaves the challenge of understanding the shear-dominated regions near the boundary.

There is a well-established phenomenological theory of the turbulent flow in the neighborhood of a flat surface.  This is presented in Landau-Lifhsitz \cite{landau}, and developments until 1989 have been reviewed by Sreenivasan \cite{sreeni89}.  This theory applies to the atmospheric surface layer, which is the source of  much of the experimental data cited in this Resource Letter. Using the language of this flow, consider the properties of turbulence at a height $z$ above a flat ground.  I will not consider any affects of roughness of the surface.  There is no natural geometrically-defined length scale for this flow.  The effective Reynolds number at height $z$ is just $Re(z)=U(z)z/\nu$, and increases with height.  The conventional theory \cite{landau} \cite{sreeni89} suggests that there is a viscous sub-layer close to the surface followed by a near-wall region in which the mean velocity $U(z)$ varies logarithmically with height $z$.  Recent high-precision measurements at Princeton \cite{zagarola97} over a wide range of  $Re$ support this logarithmic behavior. A recent theory of incomplete similarity by Barenblatt and Chorin \cite{barenblatt98} suggests, however, that a Reynolds number dependent power law is more appropriate than a logarithmic behavior.  This controversy continues \cite{smits98}, and I leave it to the reader to follow its continuing development.

\section*{CLOSING REMARKS}

Turbulence is often called the last major unsolved problem of classical physics.  In an earlier article for a general audience \cite{unsolved} I suggested that turbulence remains ``unsolved" in the sense that a clear physical understanding of the observed phenomena does not exist.  Despite major advances in computation and experiment, and theoretical advances on certain idealized model problems peripherally related to turbulence, the basic situation remains unchanged.  It is tempting to make a comparison with the equilibrium statistical mechanics of fluids.  At one time both the statistical theory of liquids and the theory of the critical point were considered unsolved problems.  The critical point had a theoretical breakthrough in the renormalization group.  The theory of liquids gradually came to be considered solved as direct computation by Monte Carlo or molecular dynamics was able to reproduce the most important observed results.  Despite the universal scaling behavior reminiscent of the scaling near the critical point, perhaps turbulence will take the path of the theory of liquids.  When we can compute enough phenomena in agreement with experiment, maybe we will decide that the problem is no longer ``unsolved."

\pagebreak

\end{document}